# Ferromagnetism in nanoscale $BiFeO_3$

R. Mazumder, P. Sujatha Devi, Dipten Bhattacharya*, P. Choudhury, and A. Sen
*Sensor and Actuator Section, Central Glass and Ceramic Research Institute, Kolkata 700032, India*

M. Raja
*Defense Metallurgical Research Laboratory, Hyderabad 500058, India*

A remarkably high saturation magnetization of ~$0.4\mu_B$/Fe along with room temperature ferromagnetic hysteresis loop has been observed in nanoscale (4-40 nm) multiferroic $BiFeO_3$ which in bulk form exhibits weak magnetization (~$0.02\mu_B$/Fe) and an antiferromagnetic order. The magnetic hysteresis loops, however, exhibit exchange bias as well as vertical asymmetry which could be because of spin pinning at the boundaries between ferromagnetic and antiferromagnetic domains. Interestingly, like in bulk $BiFeO_3$, both the calorimetric and dielectric permittivity data in nanoscale $BiFeO_3$ exhibit characteristic features at the magnetic transition point. These features establish formation of a true ferromagnetic-ferroelectric system with a coupling between the respective order parameters in nanoscale $BiFeO_3$.

*Corresponding author; dipten@cgcri.res.in

Although there has been a renewed interest recently in the area of multiferroics following the observation of a very strong interplay between magnetization (**M**) and electrical polarization (**P**) in perovskite $TbMnO_3$, $DyMnO_3$ and related $RMn_2O_5$ (R = Tb, Dy, Ho, Y etc.) systems, the search for a ferroelectric-ferromagnetic system with a strong coupling between the electric and magnetic order parameters at room temperature is still remaining futile. As of now, three different genres of multiferroic systems could be identified: (i) systems where magnetism and ferroelectricity originate in different sublattices[1] – e.g., in $BiFeO_3$ – where Bi-O orbital hybridization (or covalency) due to Bi $6s^2$ lone pair is responsible for the ferroelectric instability while Fe-O-Fe antisymmetric Dzyaloshinskii-Moriya (DM) exchange gives rise to a complicated magnetic order; (ii) systems where incommensurate spiral magnetic structure breaks down the spatial inversion symmetry[2] and thereby gives rise to ferroelectricity – e.g., in $TbMnO_3$; and (iii) systems where elastic interaction at the interface of ferroelectric-magnetic superlattice structure governs the multiferroicity – e.g., in $BaTiO_3$-$CoFe_2O_4$ multilayers.[3] In fact, encouraging developments in the latter two fields are responsible for the recent resurgence in activities in the multiferroics. Yet none of these systems could meet the criteria, outlined above, important for practical applications.

In this backdrop, the improvement in the magnetization of $BiFeO_3$ assumes importance because such improvement can help in utilizing the room temperature multiferroicity of $BiFeO_3$ ($T_C$ ~1103 K, $T_N$ ~643 K) for practical applications. In spite of room temperature multiferroicity, bulk $BiFeO_3$ suffers from poor magnetization ($0.02\mu_B$/Fe) and inhomogeneity which gives rise to leakage. We report here that a remarkably high saturation magnetization ($M_s$ ~0.4$\mu_B$/Fe) could be observed in nanoscale $BiFeO_3$ prepared by a solution chemistry route. Interestingly, even this ferromagnetic $BiFeO_3$ exhibits characteristic features in calorimetric and dielectric properties around the magnetic transition temperature ($T^*$) highlighting useful multiferroic behavior. The magnetic structure of bulk $BiFeO_3$ is complicated[4] – on the canted antiferromagnetic order between two successive (111) ferromagnetic planes, a helical order with rotation in spin direction is superposed with a periodicity ~620 Å. This, in turn, reduces the overall magnetization of $BiFeO_3$. Therefore, suppression of the helical order might give rise to higher magnetization. It has been pointed out earlier that decrease in particle size below the periodicity of the helical order can give rise to suppression of the helical order.[5]

The bulk and nano-sized powders of $BiFeO_3$ have been prepared by several solution chemistry routes – (i) co-precipitation; (ii) co-precipitation within different sol templates; (iii) auto-combustion synthesis (glycine or citrate gel); and (iv) sonochemical. Most of these techniques have been described in detail in our earlier papers.[6-8] Here we report the results of magnetic and dielectric measurements on these nanoscale powder. For the present investigation we use mostly the powder prepared by glycine combustion synthesis process (with glycine : nitrate ratio ~0.1).

The as-prepared powder is calcined at 300-700$^o$C for 4-6h in air. The calcination temperature is varied systematically in order to control the particle size of the powder. The particle size is varied over 4-40 nm. In addition, heat-treatment in flowing oxygen has also been employed in order to verify the role of excess oxygen in governing the structure and magnetic property in nanoscale $BiFeO_3$. The x-ray diffraction (XRD) patterns for all the cases have been studied at room temperature. The particle morphology and local crystallographic structure have been studied by transmission electron microscopy (TEM) and high resolution transmission electron microscopy (HRTEM), respectively. The magnetic hysteresis loops over a field range ~1.5T have been measured at room temperature for all these powders. The calorimetric and dielectric properties have also been measured across the magnetic transition point ($T^*$). The heat-treatment temperature and time, particle size, and saturation magnetization values corresponding to different samples (S1, S2, S3, and S4) are given in Table-I.

The XRD pattern for nanoscale $BiFeO_3$ (S3) is shown in Fig. 1. The data are taken with step size $0.017^o$ ($2\theta$) and step time 25s by using an advanced system of accelerator detector array. Such a detector system improves the resolution beyond what is normally achieved. This helps in improving the peak profile and identifying the phases accurately. The patterns collected thus for both the bulk and nanoscale $BiFeO_3$ have been refined by Fullprof (ver 2.3, 2003) using space group R3c. The lattice parameters, structure parameters such as bond angle, bond length etc. as well as the microstrain in the particles have been estimated and are shown in Table-II. It is quite clear that the lattice strain in nanoscale particles is nearly twice as large. Minor impurities such as $Bi_2Fe_4O_9$ (file: 20-0836) and $Bi_{24}Fe_2O_{39}$ (file: 42-0201) are found to be present in nanoscale system. The quantitative estimation shows the total concentration of impurities to be <3%. However, none of these impurities are ferromagnetic at room temperature.[9] Therefore, the room temperature ferromagnetic property observed here cannot result from any one of them. It has been pointed out earlier[10] that the presence of cubic $\gamma$-$Fe_2O_3$ impurity is responsible for higher saturation magnetization in many cases as cubic $\gamma$-$Fe_2O_3$ exhibits ferrimagnetic order with $T_C$ ~850 K. In our case, of course, such a phase is clearly absent.

In Fig. 2, we show the representative TEM and HRTEM photographs of the $BiFeO_3$ nano-particles (S3). The particles are found to be essentially multidomain with presence of interfaces between two phases – ferromagnetic (oxygen-deficient) and antiferromagnetic (stoichiometric) $BiFeO_3$. The deterioration of the room temperature ferromagnetism, observed in samples annealed under oxygen at 450°C for 6h, shows that the oxygen deficiency has a role to play for ferromagnetism. The lattice fringes in HRTEM photographs are identified to be (110), (202), (024) planes of $BiFeO_3$ phase across all the domains. The strong spin pinning at the interfaces between the ferromagnetic and antiferromagnetic phases gives rise to an exchange bias ($H_{eb}$) as well as an asymmetry in magnetization ($\Delta M$).[11]

The magnetic hysteresis loops have been measured over ±1.5T at room temperature (Fig. 3). The right y-scale of Fig. 3 depicts the magnetization of the bulk $BiFeO_3$ under identical condition. With the increase in particle size the saturation magnetization ($M_s$) decreases (Fig. 3 inset). However, over the entire range of the particle size, the ferromagnetism is retained. Both the $M_s$ and $\Delta M$ scale nearly identical patterns with particle size (d). It appears, therefore, that the interface area also decreases with the increase in particle size. The exchange bias field $H_{eb}$ [= $(H_{c1}-H_{c2})/2$; $H_{c1}$ and $H_{c2}$ are the negative and positive coercive fields, respective] varies within 110-275 Oe for particles of different sizes. Finite coercivity and $H_{eb}$ even at room temperature rule out the possibility of superparamagnetism in nanoscale $BiFeO_3$. Instead, it confirms the ferromagnetic order as well as spin pinning at the ferromagnetic-antiferromagnetic interfaces.

In Fig. 4, we show the calorimetric trace along with the representative dielectric permittivity versus temperature plot across the magnetic transition point $T^*$ for the sample S4. The dielectric property has been measured by compacting the nanoscale powder and heat-treating the samples under a moderately high temperature (~450$^o$C). The anomaly at the magnetic transition point $T^*$ is conspicuous in both these plots. It is to be noted that $T^*$ is nearly 20 K lower than $T_N$ (~653 K) of the bulk system. Of course, there is a slight mismatch between $T^*$ identified in calorimetric (~633 K) and dielectric data (~624 K). Moreover, a transition zone ($\Delta T^* = T^*-T^*_{onset}$ = 80 K) is apparent in the dielectric data. This could be because of a broader transition process expected in the nanoscale system. Further investigation is needed in order to understand the transition dynamics. However, the anomaly in the dielectric permittivity shows that the ferromagnetic phase of $BiFeO_3$ is coupled to the electric polarization which is essential for a true multiferroic system. *The observations of room temperature ferromagnetism and the coupling between ferromagnetic and ferroelectric order parameters are the central results of this paper*.

There could be combination of three factors behind the improvement in the magnetization in nanoscale particles: (i) suppression of helical order, i.e., incomplete rotation of the spins along the direction of the wave vector, (ii) increase in spin canting due to lattice strain which gives rise to weak ferromagnetism, and (iii) oxygen deficiency. The enhanced canting angle is found to have given rise to higher $M_s$ in epitaxial thin films.[12] Incidentally, in our case too, we observe increase in lattice strain in finer particles. It is worthwhile to recall here that the $M_s$ is nearly 4 times higher (~0.4$\mu_B$/Fe) in the samples prepared by us compared to what is reported very recently (~0.11$\mu_B$/Fe) in strain-free particles.[5] It appears that both the suppression of helical order as well as enhanced canting give rise to even higher $M_s$ in our case. More importantly, an evidence of coupling between magnetization and polarization is also present in these nano-particles which can certainly be exploited in nanoscale devices based on multiferroic $BiFeO_3$.

In summary, we show here that nanoscale $BiFeO_3$ depict quite high saturation magnetization as well as genuine ferromagnetic behavior with finite coercivity at room temperature. Interestingly, such a system retains the coupling between magnetization and electrical polarization and hence could prove to be quite useful for developing nanoscale multiferroic devices based on $BiFeO_3$.

We acknowledge helpful discussion with J. Ghosh on X-ray diffraction data. This work is supported by CSIR networked program Custom-Tailored Special Materials (CMM 0022). One of the authors (R.M.) acknowledges support in the form of Senior Research Fellowship (SRF) of CSIR.

[1]See, for example, H. Schmidt, Ferroelectrics **162**, 317 (1994).

[2]See, for example, S.-W. Cheong and M. Mostovoy, Nature Mater. **6**, 13 (2007).

[3]See, for example, R. Ramesh and N.A. Spaldin, Nature Mater. **6**, 21 (2007).

[4]I. Sosnowska, T. Paterlin-Neumaier, and E. Steichele, J. Phys. C **15**, 4835 (1982).

[5]T.-J. Park, G.C. Papaefthymiou, A.J. Viescas, A.R. Moodenbough, and S.S. Wong, Nano Lett. **7**, 766 (2007).

[6]S. Ghosh, S. Dasgupta, A. Sen, and H.S. Maiti, J. Am. Ceram. Soc. **88**, 1349 (2005).

[7]S. Ghosh, S. Dasgupta, A. Sen, and H.S. Maiti, Mater. Res. Bull. **40**, 2073 (2005).

[8]R. Mazumder, S. Ghosh, P. Mondal, D. Bhattacharya, S. Dasgupta, N. Das, A. Sen, A.K. Tyagi, M. Sivakumar, T. Takami, and H. Ikuta, J. Appl. Phys. **100**, 033908 (2006).

[9]N. Shamir, E. Gurewitz, and H. Shaked, Acta Cryst. A**34**, 662 (1978).

[10]H. Bea, M. Bibes, S. Fusil, K. Bouzehouane, E. Jacquet, K. Rode, P. Bencock, and A. Barthèlémy, Phys. Rev. B **74**, 020101(R) (2006).

[11]See, for example, J. Nogues and I.K. Schuller, J. Magn. Magn. Mater. **192**, 203 (1999) ; See also, A. Mumtaz, K. Maaz, B. Janjua, and S.K. Hasanain, arxiv.org : nlin/060427 (2006).

[12]J. Wang *et al*., Science **299**, 1719 (2003).

Table-I. Relevant parameters of the nanoscale $BiFeO_3$ samples

| Sample name | Heat-treatment schedule | Particle Size (nm) | $M_s$ ($\mu_B$/Fe) |
|---|---|---|---|
| S1 | Untreated | 5 | 0.41 |
| S2 | $300^0$C/6h | 15 | 0.27 |
| S3 | $450^0$C/6h | 25 | 0.13 |
| S4 | $450^0$C/48h | 40 | 0.09 |

Table-II. Relevant parameters from Rietveld refinement XRD pattern of bulk and nanoscale $BiFeO_3$ (S3); space group R3c.

| Lattice Parameters | Atom | Site | Atom Coordinates x | y | z | $R_p$ | $R_{wp}$ | $\chi^2$ | Bond Length | Bond Angle | Micro-strain (%) |
|---|---|---|---|---|---|---|---|---|---|---|---|
| **$BiFeO_3$ (bulk)** | | | | | | | | | | | |
| a = 5.578 Å | Bi | 6a | 0 | 0 | 0 | | | | Bi-O 2.309 Å | Fe-O-Fe 154.05° | 0.015 |
| c = 13.868 Å | Fe | 6a | 0 | 0 | 0.2198 | 13.6 | 22.6 | 2.17 | Fe-O 1.949 Å | O-Bi-O 73.88° | |
| | O | 18b | 0.4346 | 0.0121 | -0.0468 | | | | Fe-O 2.118 Å | | |
| **$BiFeO_3$ (nanoscale)** | | | | | | | | | | | |
| a = 5.573 Å | Bi | 6a | 0 | 0 | 0 | | | | Bi-O 2.13 Å | Fe-O-Fe 152.76° | 0.029 |
| c = 13.849 Å | Fe | 6a | 0 | 0 | 0.22324 | 4.91 | 6.68 | 9.51 | Fe-O 1.804 Å | O-Bi-O 78.93° | |
| | O | 18b | 0.4715 | 0.0119 | -0.0622 | | | | Fe-O 2.269 Å | | |

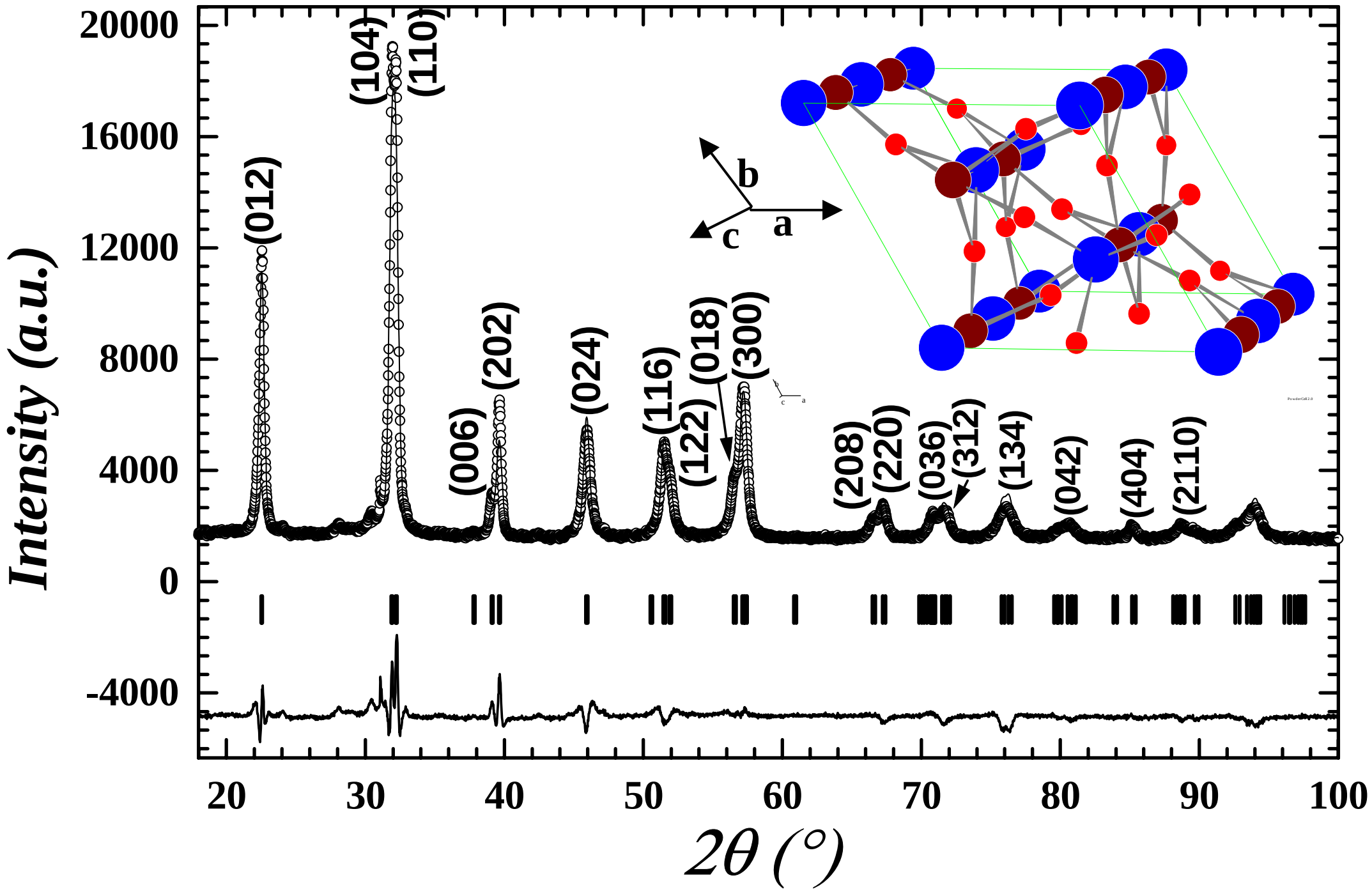

Fig. 1. (color online). X-ray diffraction patterns for nanoscale $BiFeO_3$ (S3) refined by Fullprof (ver 2.3, 2003). Insets show the cells with big, medium and small spheres correspond to Bi, Fe, and O, respectively.

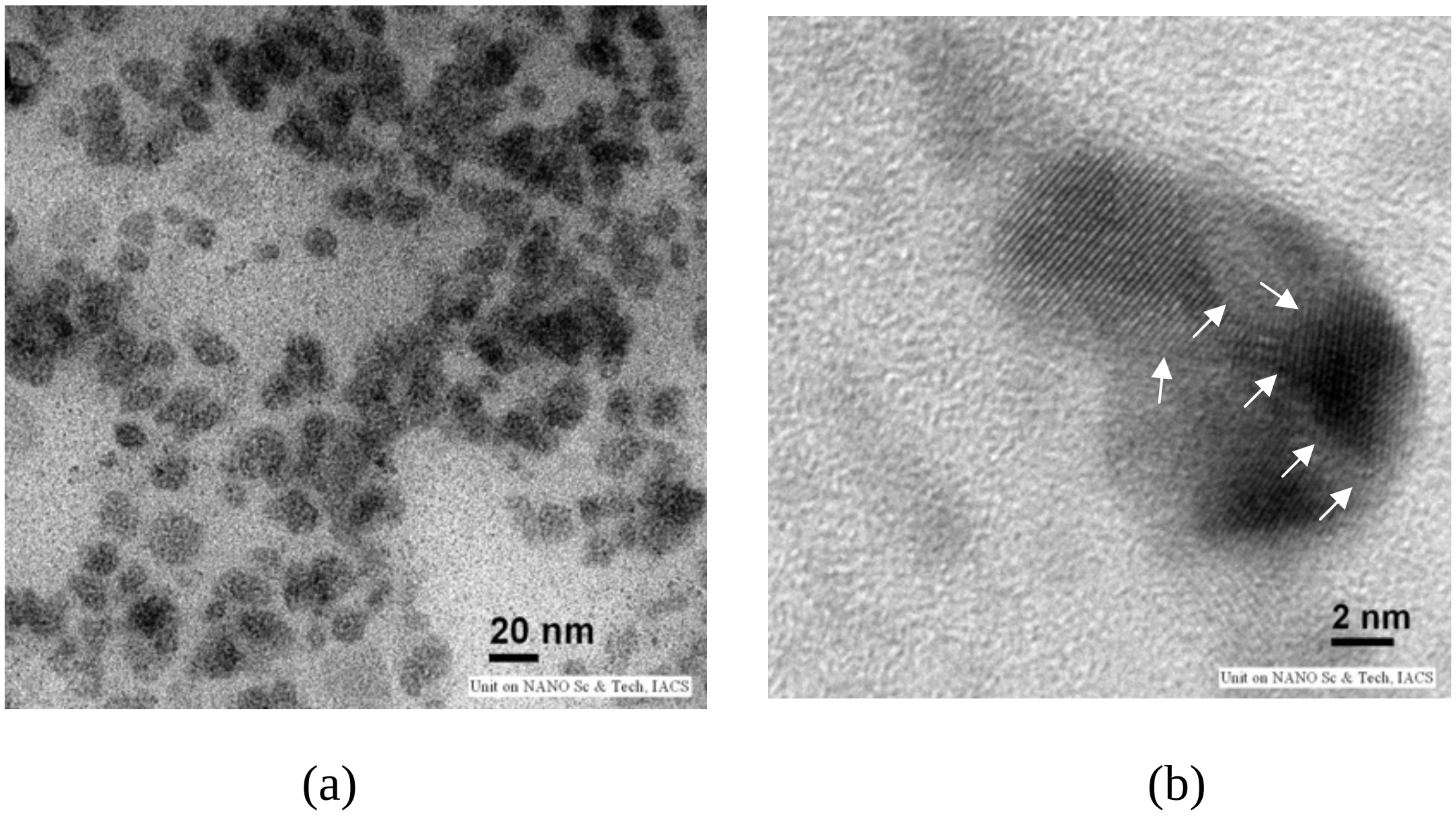

Fig. 2. (a) TEM, (b) HRTEM photographs of nanoscale $BiFeO_3$ (S3). The interfaces are shown by arrows.

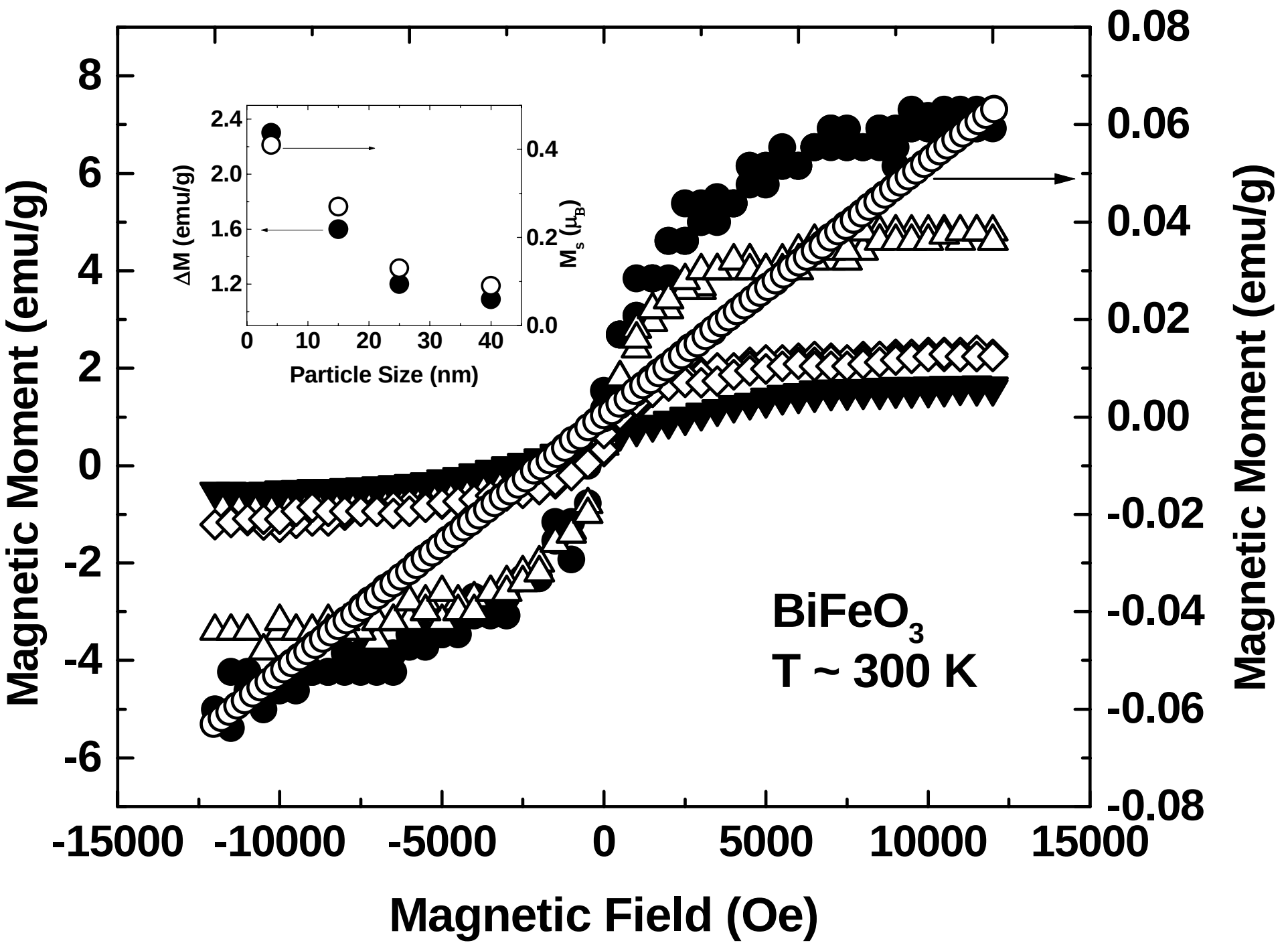

Fig. 3. Magnetic hysteresis loops for nanoscale (4-40 nm) and bulk samples (open circle). Inset: the $M_s$ (open circle) and $\Delta M$ (solid circle) versus particle size (d) patterns are shown.

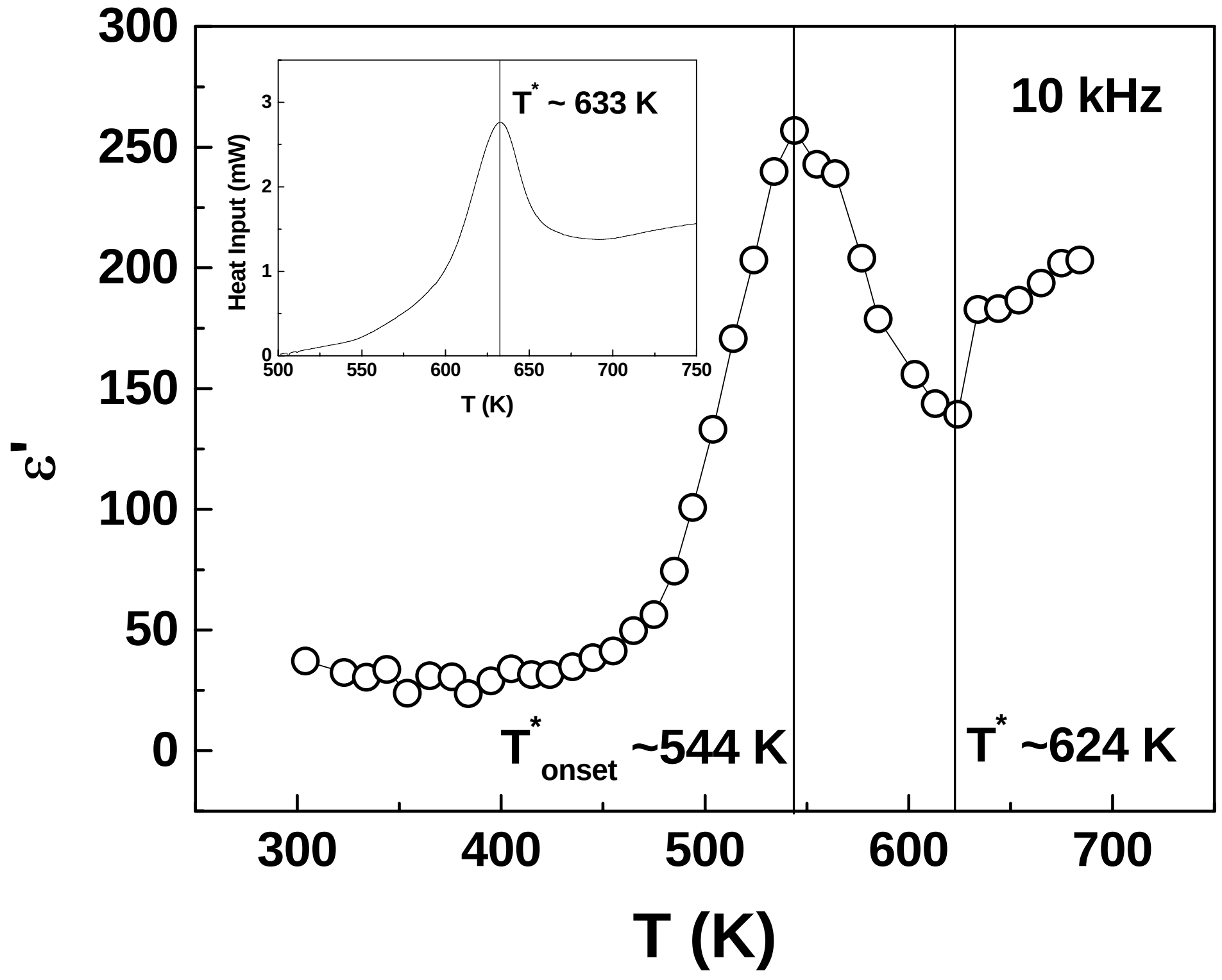

Fig. 4. Real part of the dielectric permittivity ε'(ω,T) versus temperature plot across the magnetic transition points $T^*_{onset}$ and $T^*$ for nanoscale $BiFeO_3$ (S4). Inset shows the corresponding DSC thermogram.